# Actionable Organizational States and Decision Complexity

Jan van de Poll

Transparency Lab Research Institute

## Abstract

Organizational performance indicators become useful for intervention when they can be connected to changes that management can specify, evaluate, and implement. This paper develops an organizational decision framework around a common description of practices and their permitted transitions. A1, A2, and A3 identify operational states; transition records describe the work and resources required to move between them. The resulting representation defines a conditional opportunity space and preserves the connection between a measured baseline, evaluated alternatives, and selected actions. An economic model distinguishes the value lost through incomplete information from the value lost through restricted evaluation, and connects signed practice effects to the performance decomposition in the companion research. Information and evaluation can be complements when valuable changes must be considered jointly, but substitutes when broader search reveals robust alternatives. An education case supplies 36 questionnaire items, 13 team profiles, three managers' ambitions, and indicative transition costs. A separate hypothetical six-practice model yields 389 budget-feasible configurations and permits exact Pareto and scalar benchmarks. A reference genetic algorithm approximates trade-offs without consistently outperforming random search. A selected configuration is translated into six specified transitions costing EUR 49,000 under the scenario assumptions. The contribution is an explicit, testable account of how an organizational representation can support choice and implementation. Empirical evaluation must establish whether it preserves the distinctions that matter for action and improves decisions enough to cover its construction, use, and implementation costs.

## 1 Introduction

A college seeks lower teacher absenteeism, higher diploma completion, and stronger student exit performance. These indicators identify desired outcomes. Management must still decide what to change: how attendance is recorded, how student absence is followed up, how examination results are processed, or how ICT assistance is provided. A useful decision method must connect these operational choices to their costs and consequences across the outcomes that the college values.

This paper examines how that connection can be made explicit. Each organizational practice is described by a small set of states, labeled A1, A2, and A3 in the application. A state identifies an operational condition; a transition identifies the work needed to change it. The same descriptions are used to assess the baseline, construct feasible alternatives, estimate consequences, and specify the selected work. Management can then inspect both the changes recommended and the assumptions under which those changes are attractive.

The central problem is the construction and use of a decision representation. A KPI dashboard leaves many possible interventions unspecified. An optimization model requires an action set, constraints, and consequences. An implementation plan requires tasks whose completion can be assessed. If these descriptions do not agree, a mathematically attractive recommendation may refer to a change that the organization cannot make, or the work actually performed may differ from the intervention evaluated. The framework makes those connections explicit and open to verification.

Interdependence gives this problem economic significance. One change can help one outcome while harming another, and changes that are unattractive separately can become valuable together. Shared resources and institutional dependencies further constrain the combinations available. With 36 three-state practices, the nominal space contains approximately $1.50 \times 10^{17}$ terminal profiles. This count makes the scope explicit. The difficulty of choosing depends on the interactions, constraints, uncertainty, and available decision procedures; it cannot be inferred from the count alone.

The research question is when a common representation of actionable states, combined with credible information and affordable evaluation, improves organizational choice enough to justify its costs. The paper separates possible profiles from feasible plans and distinguishes both from the value of improvement. It also separates finding an attractive plan under a model from establishing the effects of implementing that plan.

The argument continues two companion papers. Van de Poll (2026a) examines signed cross-effects, cancellation, and avoidable structural inefficiency. Van de Poll (2026b) explains how incomplete recognition and internalization of cross-unit burdens can sustain organizational layering. Here the focus moves to selection: which feasible changes can a decision process identify, how does information affect their evaluation, and how can a selected configuration be translated into specified work? The performance expression $R = E\,M(1 - C)$ enters the decision objective with its sign, joint effects, and implementation charges retained.

The contribution is a framework linking organizational measurement to constrained choice and implementation, together with an economic account of the information and evaluation it requires. The individual tools are established. Discrete states, Pareto comparison, genetic search, and costed action recommendations already have substantial literatures. The question advanced here is whether using the same operational descriptions throughout the organizational decision improves what can be evaluated and faithfully implemented. The economic results identify conditions under which that mechanism has value; the simulations demonstrate how to examine it.

The Virtual Data Scientist is the intended application of this framework. It assists with defining states, evaluating alternatives, presenting trade-offs, and specifying the transitions implied by a choice. Management supplies priorities and decision authority. Credible intervention evidence supplies the connection from operational changes to outcomes. The capability is therefore evaluated through the quality and cost of decisions and their implementation.

The education case provides the state vocabulary and indicative transition inputs. Its 13 team profiles cannot establish a flexible intervention model for 36 practices. A separate, disclosed six-practice simulation therefore makes the process reproducible without treating hypothetical outcomes as new observations. Exact enumeration provides benchmarks for both the search and the resulting action specification. Sections 2 and 3 develop the context and representation, Sections 4 and 5 the economics and decision procedure, Sections 6 and 7 the case and illustrations, and Section 8 the empirical tests. Section 9 concludes.

## 2 Relation to organizational economics and management research

### 2.1 Information and the resources required to decide

Bounded rationality makes the resources used in reaching a decision part of the decision problem (Simon, 1955). Rational inattention examines the value of acquiring information when attention is costly (Sims, 2003). Computational rationality extends the resource question to the procedures used to reason and act (Gershman, Horvitz, and Tenenbaum, 2015). These perspectives motivate the distinction between uncertainty about consequences and restrictions on the alternatives a procedure can evaluate.

Organizational economics connects knowledge and information processing to the allocation of tasks and authority. Garicano (2000) models the organization of knowledge in production. Dessein, Galeotti, and Santos (2016) show how scarce attention and adaptive decisions interact in determining organizational focus. The present paper builds on this relationship between attention and adaptation. It makes the available interventions explicit through described states, permitted transitions, and resource constraints, then considers the value of expanding the procedure used to compare them.

This distinction matters for managerial interpretation. Limited visibility is not simply an insufficient number of indicators. Information is valuable when it can alter a consequential choice. Evaluation capability concerns which alternatives can be considered and compared with that information. The framework allows the two capabilities to complement or substitute for one another. It does not assume that more information and more computation always produce a positive net return.

### 2.2 Interdependence and the difficulty of organizational choice

Interdependent practices can create organizational landscapes in which local changes do not reveal the value of coordinated changes (Levinthal, 1997). A state-based representation makes the candidate configurations countable, but countability alone does not establish computational or cognitive difficulty. A problem with many alternatives may have a simple decomposition or a strong dominance rule. A smaller problem can remain demanding when consequences are uncertain and trade-offs are difficult to evaluate.

Research on the complexity of rules also cautions against treating one numerical size measure as a complete account of human difficulty (Oprea, 2020). The empirical question here is whether managers fail to identify valuable feasible combinations under specified information, time, and resource conditions, and whether an analytical procedure improves those choices after its costs. The mathematical model identifies mechanisms to test. It does not derive a universal limit on managerial ability from the number of response categories.

### 2.3 Prescriptive analytics and actionable recourse

Bertsimas and Kallus (2020) connect predictive information to decisions and evaluate prescriptions relative to uninformed and perfect-information benchmarks. Elmachtoub and Grigas (2022) develop prediction methods whose loss reflects the downstream optimization problem. These contributions establish why predictive fit and decision value must be distinguished. An estimate matters here through its effect on the ranking and feasibility of organizational alternatives.

Actionable recourse provides a particularly close precedent. Ustun, Spangher, and Liu (2019) formulate feasible, costed input changes that alter a model’s decision. Karimi, Schölkopf, and Valera (2021) distinguish changes in predictive inputs from interventions in a causal system. The present paper applies this action-oriented perspective to collective organizational choices involving several practices, outcomes, and implementation resources. Its state descriptions must support a credible intervention interpretation before model outputs can justify real changes.

Multiobjective evolutionary optimization supplies established tools for approximating sets of trade-offs. NSGA-II combines nondominated sorting, diversity preservation, and elitist selection (Deb et al., 2002). These mechanisms are useful candidates for searching a discrete organizational space. The paper uses them as supporting computational methods. It makes no claim that applying a genetic algorithm to an organizational questionnaire is, by itself, a new optimization contribution.

### 2.4 The contribution and the claims to test

The proposed contribution concerns the construction of an organizational action set and its consistent use through implementation. A state must retain the distinctions that affect the available transitions, resource requirements, and modeled consequences. A selected configuration must map back to a permitted path from the assessed baseline. These requirements turn a questionnaire category into a candidate decision primitive.

The economic mechanism operates through the alternatives a bounded procedure can generate and compare, and through the resources consumed in doing so. Describing a coordinated change can make information about its consequences useful. Revealing an affordable robust alternative can instead reduce the value of further information. Lower interpretation and translation costs could improve the return to an analytical procedure, but those cost reductions require measurement.

The information-loss decomposition and the cancellation bridge are benchmark identities, and the approximation bound uses a standard argument. The paper's theoretical examples establish conditional mechanisms, rather than a new general theorem about managerial incapacity. Its distinctive claim is that an explicit organizational state and transition representation can connect these mechanisms to verifiable choices and implementation records. The empirical program tests that claim separately from predictive accuracy and algorithm performance.

## 3 Organizational states and feasible improvement plans

### 3.1 States as decision primitives

For practice $j$, let $\mathcal{A}_j$ contain its admissible states. In the education application, A1, A2, and A3 distinguish described ways of working. Attendance registration, for example, may use paper records, entry in a central system, or automatic registration linked to that system. The description gives the state its meaning. The label allows the same condition to be identified throughout the decision process.

A usable primitive specifies the organizational unit, reference period, state description, and evidence required to classify the condition. Its transition record specifies what must change, the resources required, and dependencies on other practices or decision-makers. These records distinguish a reported condition from a controllable choice. A college can influence the provision of ICT support, but cannot directly set a desired absence rate in the same way.

The state numbering does not establish equal distances, equal effects, or universal desirability of A3. A more elaborate arrangement may add cost or impose burdens elsewhere, leaving A1 in an efficient plan. Three states are a compact design choice; other practices can require a different number. The relevant question is whether the descriptions preserve distinctions needed to choose and implement changes at acceptable cost.

### 3.2 Targets and transition paths

Let $s = (s_1, \dots, s_{n_P})$ denote a terminal state profile, and let $\rho$ denote a permitted transition path from the incumbent profile $s^0$ to $s$. An organizational plan is $x = (s, \rho)$. Write $\Gamma(x) = \Gamma(\rho)$ for its euro expenditure. The available monetary budget is $B$, and $g$ denotes the organizational design. The set $\Omega_\rho(g)$ contains paths consistent with the relevant authority, technology, sequencing, staffing, and timing restrictions. The catalogue and admissible paths are finite over the planning horizon. The feasible plans are

$$\begin{gathered} \mathcal{X}_B(g) = \{(s,\rho)\colon\ s_j \in \mathcal{A}_j \text{ for all } j, \\ \rho\colon s^0 \rightsquigarrow s, \quad \rho \in \Omega_\rho(g), \quad \Gamma(\rho) \le B\}. \end{gathered} \tag{1}$$

This definition makes a target and the means of reaching it jointly relevant. Several paths can lead to the same terminal state while differing in cost, delay, or disruption. If such differences affect outcomes, the paths are distinct plans. When each profile has a specified path and path-independent consequences, the state vector alone can index a plan. The numerical illustration uses this simpler case: every practice begins at A1, adjacent transitions are cumulative, and each target has one specified path.

Adjacent costs cannot automatically be added in every application. Shared infrastructure can lower the cost of a bundle; conflicting interventions can make a bundle infeasible; and maintenance expenditure can continue after implementation. The transition catalogue must represent these features when they matter. An apparent recommendation to a team may also require authority at institution level. For example, school-wide attendance infrastructure cannot be treated as independently available to each team without specifying the institutional decision.

### 3.3 Three meanings of the opportunity space

The nominal number of terminal profiles is $N_{\text{nom}} = \prod_{j=1}^{n_P} |\mathcal{A}_j|$. With 36 practices and three states each, this is $3^{36} = 150{,}094{,}635{,}296{,}999{,}121$, approximately $1.50 \times 10^{17}$. This exact count describes all profiles under the stated category structure. It includes configurations that may be unreachable, unaffordable, or substantively incoherent.

The feasible opportunity space is $\mathcal{X}_B(g)$. Counting it requires resolving the actual restrictions and, where relevant, alternative paths. A search algorithm's sampled configurations do not establish the exact size of that set. Enumeration or a valid counting method is needed for an exact claim; otherwise the application should report bounds or a partial count. In the six-practice illustration, the assumptions make exact counting possible.

The economic opportunity is the improvement relative to a specified incumbent or feasible comparison, using stated outcome values, costs, and uncertainty. It can be small even when the nominal space is enormous. In a separable problem without shared constraints, management may solve the decision practice by practice. Interactions, shared budgets, dependencies, and conflicting outcomes are the mechanisms that make combinations consequential. The count supplies a description of scope, while the structure and evidence determine the difficulty and value of the decision.

### 3.4 Measurement and aggregation

The questionnaire codes A1, A2, and A3 as 0, 5, and 10. These numbers support reporting but do not establish an interval scale of capability. Treating them as numeric model inputs is an assumption to examine. A team mean of five can mean that everyone is at A2 or that equal groups are at A1 and A3. These populations can need different interventions and incur different costs despite having the same mean.

An implementable prescription therefore requires the relevant state distribution or a verified team-level state under an explicit classification rule. The intervention, response model, and transition cost must concern compatible units. A report about an individual's experience cannot automatically identify a team-wide condition. An ambition also has a distinct role: it records a preferred change over a stated horizon, rather than an implemented transition or a demonstrated optimum.

### 3.5 Preserving the distinctions that matter for action

Compression into a few states is useful only to the extent that it preserves decision-relevant differences. Consider two detailed organizational situations assigned the same code. If their permitted next steps, resource requirements, or outcome consequences differ materially, a prescription based on that code alone is under-specified. The representation must retain the distinguishing context, divide the state, or explicitly model the uncertainty within it.

An exact reduction would require every retained code to represent situations equivalent for the decision being modeled, and every selected coded plan to have a feasible implementation with the stated consequences and costs. Under those conditions, choosing among codes preserves the modeled choice. Practical applications will often use approximations. Their adequacy must be assessed against action feasibility and decision value, rather than only agreement on questionnaire scores.

The representation also needs a defined boundary. A team may choose how it enters records while institution-wide software remains outside its authority. Recording that dependency changes the feasible set; relabeling it does not remove the restriction. Conversely, a catalogue can omit a useful intervention entirely. The opportunity space in this paper is conditional on the catalogue, organizational design, and planning horizon. Maintaining it requires resources and periodic revision.

## 4 Economic value under limited visibility and evaluation

### 4.1 Valuing a feasible plan and connecting the companion papers

Let $\theta$ denote uncertain circumstances. Each $y_k(x, \theta; g)$ is an outcome change relative to a fixed baseline and horizon, oriented so that higher is desirable. The nonnegative weight $w_k$ converts that outcome into a common value scale. The implementation charge $T(x^0, x; g)$ uses the same scale. It need not equal euro expenditure numerically. Expected plan value requires explicit treatment of risks and of outcomes excluded from the chosen objective.

The first companion paper decomposes a practice's effects by their magnitude and cancellation. To connect that accounting to the present choice problem, suppose each outcome sums the individual contributions $E_i(x)\beta_{ik}$ across practices and a joint component $h_k(x, \theta; g)$. Here $E_i(x) \geq 0$ is the intensity of a practice contribution, $\beta_{ik}$ is its signed effect on outcome $k$, and $h_k$ retains joint effects that are not included in the individual terms. Define $D_i = \sum_k w_k \beta_{ik}$, $M_i = \sum_k |w_k \beta_{ik}|$, $\sigma_i = \mathrm{sgn}(D_i)$, and $C_i = 1 - |D_i|/M_i$ when $M_i > 0$. Set $C_i = 0$ if $M_i = 0$. With $H = \sum_k w_k h_k$, the same plan value has two expressions:

$$
\begin{aligned}
U(x, \theta; g) &= \sum_{k=1}^{K} w_k y_k(x, \theta; g) - T(x^0, x; g) \\
&= \sum_i \sigma_i E_i(x) M_i (1 - C_i) + H(x, \theta; g) - T(x^0, x; g).
\end{aligned}
\tag{2}
$$

Thus the shorthand $R = E\,M(1 - C)$ represents the magnitude of an individual contribution under this interpretation. Its sign, the joint effects, and implementation costs must remain present when configurations are compared. In the companion paper's per-unit notation, $R_i = |D_i|$; multiplying by intensity gives the absolute contribution $E_i R_i$. A zero cancellation index can accompany entirely harmful effects, so reducing cancellation alone is not a valid optimization objective.

The bridge is conditional on the chosen decomposition, valuation weights, baseline, and horizon. It does not identify causal effects or show that opposing effects can be removed without cost.

When the effects depend on uncertain circumstances, the entire signed product belongs inside the expectation. Multiplying separately averaged factors would generally change the value. Appendix C gives the numerical mapping for the education illustration.

### 4.2 The integrated decision problem

A signal experiment $S$ supplies information $z$ about $\theta$. Evaluation capability $b$ restricts the implementable decision procedures to $\Pi_b(S,g)$. A procedure $\pi$ maps the available signal, and any internal randomization, to a feasible plan. Information acquisition costs $C_I(S)$ and evaluation costs $C_E(b)$ are expressed in the same value units as $U$. In an application, these charges must include the relevant costs of establishing and maintaining the state catalogue and decision procedure, assigned once to avoid double counting. For a specified catalogue, the integrated economic problem is

$$J(g) = \sup_{\substack{S,b \\ \pi\in\Pi_b(S,g)}} \{\mathbb{E}[\sum_i \sigma_i E_i(\pi(z))M_i(1-C_i) + H(\pi(z),\theta;g) - T(x^0,\pi(z);g)] - C_I(S) - C_E(b)\}. \tag{3}$$

All state-dependent terms in this expression are evaluated under the same circumstances, even where their arguments are suppressed for readability. Equation (3) values a process that selects plans. Equation (2) values a particular plan. This distinction connects the performance accounting in Article 1 to the informational and computational restrictions examined here.

The supremum is a normative benchmark over specified, implementable procedures. It does not grant costless discovery of the best algorithm or signal. A useful representation can expand the alternatives a given procedure can compare or reduce the resources it needs, without changing what is physically possible. Equation (3) values that change once the procedures and costs are specified; it does not itself solve the design of the representation. Aligned objectives isolate information and evaluation. Conflicting incentives remain a separate mechanism, as in the second companion paper.

### 4.3 Information loss and evaluation loss

Write $W(S,b;g) = \sup_{\pi\in\Pi_b(S,g)}\mathbb{E}[U(\pi(z),\theta;g)]$ for attainable expected value before information and evaluation charges. The perfect-information benchmark $W^{FI}$ chooses the best feasible plan after observing the state. The unrestricted signal-based benchmark $W(S,\infty)$ chooses the best conditional expected plan after receiving $z$. For a finite feasible set and bounded payoffs,

$$
\begin{aligned}
W^{FI} - W(S,b) &= L_I(S) + L_E(S,b),\\
L_I(S) &= W^{FI} - W(S,\infty),\\
L_E(S,b) &= W(S,\infty) - W(S,b).
\end{aligned}
\tag{4}
$$

Both losses are nonnegative when the stated procedure classes respect the information restriction. This is a benchmark decomposition, not an empirical measurement of a particular manager's shortcomings. It requires knowledge of the relevant feasible alternatives and their consequences. Implementation costs already enter each benchmark through $U$, while information and evaluation charges are considered separately through $J$.

More informative signals weakly improve unrestricted expected value in the Blackwell sense (Blackwell, 1953). The same conclusion holds for restricted procedures when they can ignore or appropriately garble the additional information without losing access to the old procedure. Larger nested procedure classes also weakly increase gross attainable value. Net value need not rise after acquisition costs. Appendix A defines a normalized value-based visibility measure and explains its relationship to existing prescriptive benchmarks.

### 4.4 When information and coordinated evaluation are jointly valuable

Consider two binary changes whose individual costs are $c > 0$. Their joint benefit is $\theta V$, where $\theta$ equals minus one or plus one with equal probability and $V > 2c$. Isolated changes therefore lose money; a coordinated change is valuable only in the favorable state. Let signal precision $e \in [0,1]$ imply probability $(1+e)/2$ of correctly identifying that state. Acquiring precision costs $\kappa e^2/2$, with $\kappa > 0$, and enabling coordinated evaluation costs $F_E \geq 0$.

When the procedure considers only isolated changes, information cannot justify adoption. With coordinated evaluation available, the gross expected value becomes $1/2\,(Ve - 2c)_+$, where $(t)_+ = \max(t, 0)$. Define

$$
\hat{e} = \min\{V/(2\kappa), 1\}, \qquad \Phi = V\hat{e}/2 - c - \kappa\hat{e}^2/2 - F_E. \tag{5}
$$

**Proposition 1.** In this two-change model, the optimal net project value is $\max(0, \Phi)$. If $\Phi > 0$, the organization acquires coordinated evaluation capability and precision $\hat{e}$. If $\Phi < 0$, it retains the incumbent without those investments. At equality, retention is an optimum.

The result gives a concrete mechanism for the proposed representation. A procedure that can describe and compare a coordinated transition makes information about its consequences potentially useful. In this example, an uninformative signal cannot justify coordinated adoption. Practical recommendations additionally require feasible transitions and sufficiently credible effect estimates, with gains exceeding total costs.

Complementarity is conditional. If expanded evaluation instead reveals a robust option that performs well across states, additional information can become less valuable. Appendix A gives finite examples of both signs and extends the model to the choice between a simple and an interdependent organizational design. These results establish conditional economic relationships rather than a general preference for organizational complexity or analytical investment.

## 5 From feasible states to choice and implementation

### 5.1 Pareto comparison with explicit objectives

When management has not fixed the relative valuation of the outcomes, the procedure can first identify trade-offs. Let $\mathbf{f}(x \mid z)$ contain the conditional expected, favorably oriented outcomes and negative expenditure. A plan dominates another when it is at least as good in every included component and strictly better in at least one. The conditional Pareto set is

$$\begin{aligned} \mathbf{f}(x \mid z) &= (\mathbb{E}[y_1(x,\theta) \mid z], \ldots, \mathbb{E}[y_K(x,\theta) \mid z], -\Gamma(x)), \\ \mathcal{P}(z) &= \{x \in \mathcal{X}_B : \nexists x' \in \mathcal{X}_B \text{ with } \mathbf{f}(x' \mid z) \succ \mathbf{f}(x \mid z)\}. \end{aligned} \tag{6}$$

The design is fixed in this notation. Costs enter the frontier when expenditure remains relevant below the budget ceiling. A frontier based on three KPIs alone can discard a cheaper plan that is preferred after costs. The simulation therefore uses four objectives: lower absence, higher completion, higher exit index, and lower euro cost. The hard budget and the preference for lower expenditure play different roles.

A Pareto set identifies alternatives that cannot be improved in all included dimensions simultaneously. It does not select a single managerial decision. Selection still requires priorities, acceptable outcome thresholds, or another explicit preference rule. A linear weighted objective is one transparent rule, but varying positive weights need not recover every Pareto-efficient point in a discrete nonconvex set. A three-dimensional display is also only a projection when expenditure is a fourth objective.

### 5.2 What a genetic algorithm contributes

A discrete genetic search encodes a candidate as the selected state for each practice. Initialization supplies feasible candidates, recombination combines parts of existing configurations, and mutation changes individual practice states. Feasibility checks reject or repair configurations that violate the stated restrictions. Nondominated sorting and diversity measures preserve useful trade-offs during selection. These operations evaluate combinations that would be burdensome to inspect manually when the relevant search is extensive.

The returned nondominated archive is generally an approximation. It can omit efficient plans and include plans dominated by feasible alternatives that were never evaluated. A search trace should record the representation, constraints, objective model, evaluation budget, random seeds, and handling of infeasible or repeated candidates. On tractable instances, exact enumeration or a suitable exact solver supplies a benchmark. On larger instances, comparison with strong baselines and available bounds is necessary before claiming optimization performance.

A genetic algorithm is one candidate method. Integer programming, dynamic programming, decomposition, local search, or exhaustive enumeration may be preferable for particular structures. The scientific proposition concerns decision value under resource constraints, so the method should be selected by its performance on the relevant problem rather than by its label. Section 7 includes a modest reference comparison and reports its unfavorable as well as favorable results.

### 5.3 Prediction error remains after search improves

Let $m(x) = \mathbb{E}[U(x,\theta) \mid z]$ be correct conditional expected value and $\widehat{m}(x)$ its estimate. Suppose $\sup_x|\widehat{m}(x) - m(x)| \leq \varepsilon$ on the feasible set and the selected plan $\hat{x}$ is within $\delta$ of the maximum estimated value. Then

$$\max_{x \in \mathcal{X}_B} m(x) - m(\hat{x}) \leq 2\varepsilon + \delta. \tag{7}$$

**Proposition 2.** Under these uniform valuation and optimization conditions, the loss relative to the best conditional expected plan is bounded by twice the valuation error plus the optimization error.

This elementary bound separates two sources of failure. A faster search can reduce $\delta$ without reducing $\varepsilon$. A uniform error bound over untried organizational configurations is demanding and cannot be inferred from an average validation error alone. An uncertified heuristic also does not supply a valid numerical value for $\delta$ merely by terminating. The bound is a diagnostic framework, with operational values available only when the assumptions can be supported.

Optimization can further select unusually optimistic estimates, creating postdecision disappointment (Smith and Winkler, 2006). Sensitivity analysis, independent validation, and cautious treatment of unsupported configurations are therefore part of the method. A predicted change from replacing a questionnaire input is an intervention claim only when the model and the evidence justify that interpretation.

### 5.4 Translating selection into an action plan

The selected configuration becomes an action specification by resolving each difference from the verified baseline into permitted transitions. Each transition retains its source and target state, implementation task, resource estimate, dependencies, responsible role, and completion evidence. The resulting record can be checked against the plan whose outcomes were evaluated.

This translation has concrete correctness conditions. The steps must start from the assessed baseline, follow permitted transitions, and end at the selected target. Their resource requirements must reconcile with the plan's cost model, including any shared costs or interactions. Predecessors and decision rights must agree with the feasibility restrictions. Passing these checks establishes consistency between the model and the specification. Successful implementation still depends on the operational facts and the work performed.

For test-result processing, selecting A3 from A1 requires establishing a digital results system and linking it to the student tracking system under the assumed adjacent path. Recording only a target score of ten does not identify these tasks. Procuring software alone may also leave the target incomplete. The action record must state how completion will be verified.

An executable plan additionally needs assigned staff, agreed service standards, and a schedule consistent with resource availability. Unresolved fields remain explicit. After implementation, the same identifiers connect completed transitions to actual expenditure and subsequent outcomes. This permits separate investigation of incomplete implementation, incorrect resource estimates, and inaccurate effect models. The education illustration checks the translation mechanically and identifies the operational details that still require decisions.

## 6 The education case and measurement boundaries

### 6.1 Available records

The supplied workbook contains 13 team profiles, three KPI columns, and 36 questionnaire items. Of 468 possible team-item entries, 461 are populated. The seven missing entries occur in one team. The KPI labels concern teacher absenteeism, diploma completion, and students' exit performance. Exact observation periods, denominators, and the units of the exit-score measure are not fully documented in the supplied materials.

The team questionnaire values are averages of respondent codes 0, 5, and 10. Respondents report the current situation and their near-term ambition, with the latter intended to identify priorities rather than an unconstrained ideal. The workbook also supplies the six-month ambition profiles of three managers. These are three managers from the participating group, not a controlled comparison of independently developed plans under an established common baseline, information set, objective, and budget.

The cost sheet gives indicative euro costs per team for A1-to-A2 and A2-to-A3 transitions, along with person-hours and nominal durations. Whether monetary figures already include the listed hours is unresolved. The simulation uses the monetary figures as stated and reports hours separately. It does not add an assumed wage bill to those costs or treat the resulting hours as a demonstrated implementation capacity.

### 6.2 Verifiability is a property to establish

The survey format follows the intention to ask about recognizable facts and behavior while capturing current states and ambitions (Van de Poll, 2021). It offers a useful starting point for specifying interventions. However, the descriptions in the supplied workbook vary in their observability. Some refer to identifiable systems; others use terms such as "sometimes," "immediate," or "much time." The daily-work-location options also overlap at five locations. These features require clarification before the full catalogue can be treated as mutually exclusive, verifiable states.

The proposed method therefore imposes a stronger requirement than assigning codes to questionnaire responses. Each retained state needs an operational definition, and the relevant baseline requires verification. For ICT support, "immediate help whenever needed" should be replaced by an agreed service threshold and a defined population of requests. That revision would require reassessing its costs and outcomes. The current illustration preserves the source description and flags the unresolved threshold rather than silently inventing a validated state.

The survey's focus on behavior does not eliminate self-report error or establish the validity of aggregation. A statement can be verifiable without having been independently verified. Nor does the absence of attitudinal normalization resolve the scale issue: equal numerical gaps between codes remain a modeling convention, and different KPI units still require explicit valuations when combined. A methodological comparison of response formats cannot, on its own, validate this particular intervention model.

### 6.3 Available model evidence

A supplied model-comparison table selects Extra Trees for absenteeism and reports $R^2 = 0.014$ and mean absolute error 0.042. The other displayed models have negative $R^2$. Under the usual definition, an $R^2$ close to zero indicates performance close to the evaluation sample's mean-prediction reference. If the MAE uses the original proportion scale, it corresponds to 4.2 percentage points. The validation protocol and uncertainty of these estimates are not documented in that table.

These results do not establish useful intervention rankings. Thirteen team-level outcomes remain 13 observations even if the questionnaire means summarize many respondents. Restricted or regularized models can be fitted, but unrestricted effects of 36 variables and their interactions are not identified by this sample. A stronger predictive score would still require a separate causal argument before treating changes in inputs as effects of organizational interventions.

The case therefore supplies a vocabulary of candidate states, ambitions, outcomes, and transition requirements. It does not supply estimated response functions for the simulations, demonstrate that any manager chose poorly, or establish product effectiveness. Artificially extending the observed sample would not change that evidential status. Synthetic records can test software under stated assumptions but do not create independent observations of educational responses.

# 7 Reproducible computational illustrations

## 7.1 The economic investment threshold

Set $V = 8$, $c = 1$, and $F_E = 0.5$ in the coordinated-change model. A simple organizational design yields $v_S = 0.8$, while switching to the interdependent design costs $C_{\text{switch}} = 0.2$. The preferred design compares $v_S$ with $\max(0, \Phi) - C_{\text{switch}}$. Table 1 gives selected evaluations of the closed-form result. All quantities use hypothetical common value units.

**Table 1 Information costs and the design decision**

| Information cost κ | Chosen precision | Net project value | Value after design transition | Preferred design |
|---|---|---|---|---|
| 1 | 1.00 | 2.00 | 1.80 | Interdependent |
| 2 | 1.00 | 1.50 | 1.30 | Interdependent |
| 3 | 1.00 | 1.00 | 0.80 | Simple at tie |
| 4 | 1.00 | 0.50 | 0.30 | Simple |
| 5 | 0.80 | 0.10 | −0.10 | Simple |
| 6 | 0.00 | 0.00 | −0.20 | Simple |

The example distinguishes investment within a design from the decision to adopt that design. At $\kappa = 5$, the coordinated project justifies information and evaluation expenditures, yet switching away from the simple design remains unattractive. At $\kappa = 2$, the interdependent design is preferable after the transition charge. Lower decision costs can therefore change the preferred organizational arrangement, but the result depends on the joint benefits and the available alternative.

## 7.2 A six-practice education scenario

The second illustration selects six practices to permit exhaustive verification. Each starts at A1 and has implementation level $x_j \in \{0, 0.5, 1\}$ corresponding to A1, A2, and A3. These levels are scenario encodings, not observed team means or proven equal increments in capability. Table 2 retains the workbook question identifiers and indicative euro costs.

**Table 2 Practices and adjacent transition costs**

| Code | Item | Practice | A1 to A2 euros | A2 to A3 euros |
|---|---|---|---|---|
| $x_1$ | Q1 | Recording student attendance | 8,500 | 28,000 |
| $x_2$ | Q3 | Following up student absence | 2,500 | 16,000 |
| $x_3$ | Q12 | Processing and storing test results | 9,000 | 14,000 |
| $x_4$ | Q19 | Daily work locations | 2,000 | 4,500 |
| $x_5$ | Q31 | Availability of ICT assistance | 6,000 | 9,000 |
| $x_6$ | Q34 | Administrative work outside teaching | 3,500 | 14,000 |

A move to A3 incurs both adjacent costs. The monetary restriction is

$$\Gamma(x) = \sum_{j=1}^{6} [c_{12,j}\mathbf{1}\{x_j \geq 0.5\} + c_{23,j}\mathbf{1}\{x_j = 1\}] \leq 50{,}000. \tag{8}$$

Of $3^6 = 729$ nominal profiles, 389 satisfy the assumed budget. This is an exact count for the stated simplified catalogue. No additional staffing, institutional authority, or scheduling restrictions are imposed in the computational instance. Its feasibility claim is consequently limited to its explicit transition and monetary assumptions.

Appendix B specifies three hypothetical response functions: absence rate $a$, completion rate $d$, and exit-performance index $q$. Two independent state variables, $\theta_1$ and $\theta_2$, take values minus one and plus one with equal probability and determine the signs of two interactions. All coefficients and baseline outcomes are assumptions. The scalar preference rule used for illustration is

$$\begin{gathered} U(x,\theta) = 100\{0.08 - a(x,\theta) + d(x,\theta) - 0.65\} \\ +100\{q(x,\theta) - 0.70\} - \Gamma(x)/10{,}000. \end{gathered} \tag{9}$$

A percentage-point decrease in absence or increase in either other outcome receives one value unit, and EUR 10,000 of expenditure costs one value unit. These conversions are stated preferences. They do not establish economic equivalence among educational outcomes. The exit index belongs to the simulation and is not a normalization attributed to the observed exit-score column. The incumbent has value zero.

### 7.3 Information and evaluation effort

The information experiment varies information independently of the number of evaluated candidates. With interaction signs hidden, selection maximizes prior expected value. With signs revealed, selection maximizes value in the known state. Uniform random search evaluates the incumbent plus 10 or 50 draws, with replacement, from other feasible configurations. The higher budget retains the first ten draws. Table 3 reports 10,000 replications, integrating the four states exactly within each replication.

**Table 3 Expected value under information and evaluation restrictions**

| Information | Additional evaluations | Expected value | Monte Carlo SE | Gap to full information and exhaustive search |
|---|---|---|---|---|
| Signs hidden | 10 | 11.887 | 0.017 | 4.063 |
| Signs revealed | 10 | 12.165 | 0.016 | 3.785 |
| Signs hidden | 50 | 13.824 | 0.011 | 2.126 |
| Signs revealed | 50 | 14.254 | 0.009 | 1.696 |
| Signs hidden | Exhaustive | 15.900 | 0.000 | 0.050 |
| Signs revealed | Exhaustive | 15.950 | 0.000 | 0.000 |

The information gain rises from 0.278 at ten draws to 0.430 at fifty. The paired cross-difference is 0.152, with Monte Carlo standard error 0.0044. Information and evaluation are complementary for this change in the search budget. Under exhaustive evaluation, however, the information gain is only 0.050. A strong prior-optimal plan reduces the need to learn the interaction signs. Complementarity therefore depends on the capabilities being compared.

The values include implementation expenditure but exclude information and search charges. The feasible list is supplied to the procedure, so the experiment measures outcome evaluations rather than the cost of finding feasible candidates. A sensitivity calculation varies three budgets and two sets of three outcome weights, giving 27 settings. The estimated cross-differences range from −0.006 to 0.367. These calculations describe the assumed family and are not independent empirical replications.

### 7.4 Exact Pareto sets and a reference genetic search

The Pareto experiment keeps all four objectives explicit and reveals the interaction state. Exact enumeration produces 137, 127, 149, and 138 nondominated configurations in states $(-1,-1)$, $(-1,1)$, $(1,-1)$, and $(1,1)$, respectively. The relatively large sets illustrate that eliminating dominated choices can still leave management with substantial preference decisions.

A reference discrete genetic algorithm uses NSGA-II ranking, crowding, and elitist population selection, with uniform crossover and categorical mutation. It is a transparent implementation for this illustration, not an export of the PRAIORITIZE product. Population size is 20. The comparison uses 50 runs in each state, evaluation budgets of 60 and 200 including the incumbent, and independent uniform random search with the same evaluation budgets. Repeated candidates consume evaluations in both procedures. Appendix B records all settings.

**Table 4 Search results averaged over four states and 50 runs per state**

| Method | Evaluations | Mean distinct plans | Exact frontier recovered | Mean value gap | Runs finding scalar optimum |
|---|---|---|---|---|---|
| Random | 60 | 55.89 | 14.82% | 1.589 | 12.0% |
| Genetic | 60 | 48.97 | 15.23% | 1.863 | 13.5% |
| Random | 200 | 156.48 | 40.22% | 0.649 | 37.5% |
| Genetic | 200 | 123.45 | 39.11% | 0.813 | 37.5% |

Frontier recovery is the fraction of the state's exact Pareto configurations present in the nondominated archive of all evaluated candidates, averaged across runs. The value gap uses equation (9) to compare the best evaluated plan with its exact statewise optimum. This is a supplementary preference-based diagnostic; the genetic algorithm searches all four objectives and is not tuned to that scalar rule. The reported percentages and gaps are descriptive simulation averages rather than population estimates of managerial performance.

The genetic procedure does not consistently outperform random search in this instance. It obtains slightly greater frontier recovery at the lower budget, while random search has a smaller mean scalar gap at both budgets and greater frontier recovery at 200 evaluations. Neither procedure reliably recovers the full frontier. The illustration supports the ability to run and audit the full decision process, while offering no evidence that a genetic algorithm is generally superior. Equal outcome-evaluation budgets also do not equalize every computational cost; proposal counts are retained in the outputs.

### 7.5 From a selected configuration to specified changes

In the favorable interaction state $(1,1)$, equation (9) selects the exact optimum $(0.5,0.5,1,0,1,0)$. It changes four practices through six adjacent transitions, leaves Q19 and Q34 at A1, and costs EUR 49,000. Simulated outcomes move from absence 8.00% to 4.95%, completion 65.00% to 75.25%, and exit index 0.7000 to 0.8125. The net value is 19.65 under the stated preference rule. These are model-generated consequences, not observed educational improvements.

Table 5 compiles the selected transitions. Endpoint descriptions are concise English translations of the workbook. The hours remain indicative source inputs. Proposed owner roles and verification criteria are supplied in the reproduction package as implementation templates, without implying that staff have been assigned or the work approved.

**Table 5 Transition specification for the selected simulated plan**

| Item and step | Change in organizational state | Euros | Staff hours |
|---|---|---|---|
| Q1 A1 to A2 | Replace attendance records on paper with entry in a central system | 8,500 | 82 |
| Q3 A1 to A2 | Introduce separate follow-up actions after absence registration | 2,500 | 32 |
| Q12 A1 to A2 | Establish a digital system for test results | 9,000 | 86 |
| Q12 A2 to A3 | Link test results automatically to the student tracking system | 14,000 | 106 |
| Q31 A1 to A2 | Move from no ICT support to some support, not always timely | 6,000 | 62 |
| Q31 A2 to A3 | Move to the stated target of immediate ICT assistance when needed | 9,000 | 82 |
| Total | Six transitions across four practices | 49,000 | 450 |

For Q12 and Q31, the second transition depends on the first under the assumed path. The source's nominal durations are also exported, but adding them does not establish an elapsed schedule when work can overlap or compete for staff. The 450 hours are reported separately from euro costs to avoid an unsupported assumption about double counting.

The ICT target remains insufficiently operationalized. Before this plan could be implemented, the organization would need to define the service standard, verify available capacity, revise cost estimates as appropriate, and re-evaluate the corresponding effects. The example demonstrates traceability from selection to specified changes while making the unresolved implementation fields visible. That traceability is the practical purpose of maintaining the same primitives throughout the process.

# 8 Empirical implications and research agenda

## 8.1 Testing the representation separately from the optimizer

A controlled study should first test whether the representation improves the definition and translation of the decision. Participants can receive the same substantive facts, outcome model, costs, and restrictions in two formats: ordinary organizational descriptions or a consistent state and transition catalogue. Holding content constant distinguishes the effect of representation from the effect of providing additional information. Access to automated search can then be varied separately.

The testable expectation is fewer infeasible proposals, fewer omitted dependencies, and closer correspondence between evaluated targets and specified actions when the catalogue preserves the relevant distinctions. Decision time and catalogue-construction costs must also be recorded. The claim is weakened if benefits disappear under equal information or if classification effort exceeds the resulting gains. Tasks with ambiguous states provide a direct test of the representation's limits.

Independent assessors should classify the baseline and compare generated action records with the selected targets. Subsequent field evaluation should check whether the specified transitions are completed as intended. This separates agreement about a code, correctness of its translation, and actual implementation. A common vocabulary can improve one without improving all three.

### 8.2 Testing information and evaluation mechanisms

A separate factorial study can vary decision-relevant information and access to a specified evaluation procedure. Participants should face the same baseline, feasible interventions, objective, time allowance, and implementation budget. Controlled tasks with known consequences permit measurement of decision loss. The theory predicts complementarity when information identifies valuable coordinated changes that the restricted procedure cannot evaluate. It predicts substitution when expanded evaluation reveals an attractive option that is robust across states.

The interaction is therefore a hypothesis tied to the task structure. Varying only the number of nominal combinations would not adequately test it. Studies should also vary the strength of practice interactions, the shared constraints, and the availability of robust plans. An observed advantage from analytical assistance must be assessed after its costs and against appropriate simple rules and optimization baselines.

This design makes claims about cognitive difficulty falsifiable. Managers may use structure to choose well without exhaustive search. Evidence of difficulty requires poorer choices under specified conditions, and evidence for the proposed assistance requires an improvement under a comparable decision problem. The three managers' ambition profiles in the workbook cannot establish either claim because equivalence of their baselines, information, objectives, and budgets has not been established.

### 8.3 Establishing intervention effects

Prospective implementation evidence is the main requirement for moving from model-based prescriptions to management conclusions. Where feasible, teams can be assigned to well-defined interventions or to implementation sequences. Alternatives include credible quasi-experimental designs with documented assumptions about selection, timing, and spillovers. The intervention unit and the outcome unit must agree, and institution-wide changes may require institution-level treatment definitions.

The study should retain baseline states, response distributions, assigned and completed transitions, actual resource use, and outcome periods. Educational consequences may emerge later than the six-month ambition horizon. A change that improves completion by altering assessment standards could have different implications from one that improves learning. Outcome definitions and relevant safeguards therefore belong in the scientific specification of the decision.

Predictive evaluation should use genuinely unseen teams or later periods where the data permit. Imputation, feature construction, selection, and synthetic-data generation must occur within training partitions, preventing evaluation information from entering model development (Cawley and Talbot, 2010). Repeated splitting does not create independent teams. Synthetic experiments can identify failure modes under declared generating processes, but causal and external validation require observations beyond the present small sample.

### 8.4 Evaluating the Virtual Data Scientist

The Virtual Data Scientist implements connected analytical tasks: assessing states, estimating consequences, generating feasible alternatives, searching, eliciting priorities, specifying work, and recording implementation. Each component should be evaluated against its purpose. Better prediction can leave the selected plan unchanged, while resolving a small uncertainty near a decision threshold can change the choice substantially.

The primary economic outcome is the gain in realized or credibly estimated decision value after development, information, evaluation, implementation, and maintenance costs. Relevant secondary outcomes include feasibility, state classification reliability, translation accuracy, uncertainty calibration, and implementation fidelity. A Pareto display is an intermediate output whose value depends on the choices it supports.

Evaluation should progress from known computational problems to controlled decisions and prospective organizational interventions. Simulation reports should distinguish generating assumptions, procedures, performance measures, and Monte Carlo uncertainty (Morris, White, and Crowther, 2019). The present package supports computational examination of the mechanism. Prospective studies must establish whether its benefits persist when effects, costs, and implementation are uncertain.

### 8.5 Boundaries and the continuing research line

The model treats the feasible catalogue and distribution of consequences as specified. It does not determine the optimal state vocabulary, identify omitted interventions, or estimate causal effects. A finer representation can preserve more relevant distinctions while increasing assessment and evaluation costs. Studying that trade-off is a natural theoretical extension: the representation itself becomes an organizational choice.

The current objective also assumes agreement about what the organization values. Better recognition of a cross-unit burden need not make the responsible manager internalize it. The link to the second companion paper is therefore substantive. Decision rights and incentives would be needed to distinguish consequences that are unknown, consequences that cannot be evaluated affordably, and consequences that are known but disregarded.

A dynamic extension should include delayed effects, reversibility, maintenance, and learning from implemented changes. An intervention can change future information as well as current outcomes. Stable identifiers can preserve the history of those changes, while versioned definitions record when the meaning of a state has changed. Neither the catalogue nor the estimated response model should be assumed permanent.

The research line thus advances from accounting for signed organizational effects, through the mechanisms that sustain them, to the conditions for selecting and implementing feasible changes. Its next empirical step is a controlled test of the representation and decision procedure, followed by prospective evaluation of well-defined interventions. General management claims require that accumulated evidence.

## 9 Conclusion

A1/A2/A3 can serve as an operational vocabulary connecting an assessed organizational condition to feasible changes and the work needed to realize them. Its usefulness depends on preserving the distinctions that affect feasibility, consequences, and resources. With credible effect models, the resulting catalogue permits explicit comparison of trade-offs and translation of selected configurations into action specifications.

The economic model explains why information and evaluation must be considered together and why their relationship can change across decision problems. The signed bridge to $R = E\,M(1 - C)$ retains opposing effects, joint contributions, and implementation charges. The education simulation demonstrates exact counting, Pareto comparison, heuristic assessment, and action translation under disclosed assumptions. It also shows that a genetic algorithm need not outperform a simple search baseline.

The scientific claim to test is whether this common representation enables better feasible choices and more faithful implementation at an acceptable total cost. Establishing that result would provide a foundation for the Virtual Data Scientist as an organizational decision capability.

# Appendix A Economic derivations

## A.1 Benchmarks and decision relevant visibility

For fixed design and feasible set, the benchmark values are

$$\begin{aligned} W^{FI} &= \mathbb{E}[\max_{x \in \mathcal{X}_B} U(x,\theta)], \\ W(S,\infty) &= \mathbb{E}_z[\max_{x \in \mathcal{X}_B} \mathbb{E}[U(x,\theta) \mid z]], \\ W_0 &= \max_{x \in \mathcal{X}_B} \mathbb{E}[U(x,\theta)]. \end{aligned} \tag{A1}$$

A randomized feasible decision conditional on $z$ has value no greater than the highest conditional expected payoff. Taking expectations and the supremum over permitted procedures yields $W(S,b) \le W(S,\infty)$. The expectation of a statewise maximum is at least the maximum of expected values, yielding $W(S,\infty) \le W^{FI}$. Adding and subtracting the intermediate benchmark proves equation (4). Blackwell monotonicity for restricted procedures additionally requires closure under ignoring or garbling the signal at the relevant capability level.

When $W^{FI} > W_0$, a value-based visibility index is

$$\eta(S;g) = \frac{W(S,\infty;g) - W_0(g)}{W^{FI}(g) - W_0(g)}. \tag{A2}$$

This index lies in $[0,1]$ under the stated benchmark assumptions. It is undefined when the denominator is zero, since no positive perfect-information gain is available to normalize. It is not the fraction of organizational records observed. It is also distinct from $1 - C_i$, which measures cancellation within a practice's valued effects. The normalization is closely related to the coefficient of prescriptiveness in Bertsimas and Kallus (2020), specialized here to the best decision under a signal. No independent novelty is claimed for this ratio.

## A.2 Coordinated change and proof of Proposition 1

The two-change example uses binary $x_1, x_2$ and payoff

$$U(x_1, x_2, \theta) = \theta V x_1 x_2 - c(x_1 + x_2). \tag{A3}$$

The incumbent gives zero and either isolated change gives $-c$. At low evaluation capability, only these alternatives are considered. At high capability, the coordinated change is available. Symmetric signal accuracy $(1+e)/2$ gives posterior mean $e$ in the favorable signal and $-e$ in the unfavorable signal. Each signal occurs with probability one half. Therefore

$$W(S_e, 1) = 1/2\,(Ve - 2c)_+, \qquad W(S_e, 0) = 0. \tag{A4}$$

Where $Ve \le 2c$, investment has no positive gross value and cannot dominate retention after its nonnegative costs. On the active branch, net value is $h(e) = Ve/2 - c - \kappa e^2/2 - F_E$. This concave quadratic has maximizer $\hat{e}$ in equation (5) over $[0,1]$. If its maximum $\Phi$ is positive, the maximizing point necessarily satisfies $V\hat{e} > 2c$ and lies on the active branch. If $\Phi \leq 0$, no active-branch point improves on retention. This proves Proposition 1.

For $\kappa \geq V/2$, substitution gives $\Phi = V^2/(8\kappa) - c - F_E$. For $\kappa < V/2$, precision is saturated at one and $\Phi = V/2 - c - \kappa/2 - F_E$. The formulas agree at the boundary. Retention is chosen at a tie; redundant costless capabilities can also be optimal without changing value.

### A.3 Complementarity and a counterexample

Let the first index of $W_{ab}$ indicate low or high information and the second indicate low or high evaluation capability. Define

$$D_{IE} = W_{11} - W_{10} - W_{01} + W_{00}. \tag{A5}$$

In the coordinated-change example, comparing uninformative with perfect signals and isolated with coordinated evaluation gives $W_{00} = W_{10} = W_{01} = 0$ and $W_{11} = (V - 2c)/2$. Thus $D_{IE} > 0$.

For a counterexample, consider two equally likely states and two ordinary actions. Each pays one in its matching state and zero in the other. At low evaluation capability these are the only available actions. At high capability a robust action paying 0.6 in both states is added. Uninformative values are 0.5 and 0.6, while perfect information yields one under either capability. Hence $(W_{00}, W_{01}, W_{10}, W_{11}) = (0.5, 0.6, 1, 1)$ and $D_{IE} = -0.1$. Information and evaluation are substitutes in this finite example.

### A.4 Search coverage and valuation error

Fix a signal and correct candidate values. Suppose a pool of $N$ candidates contains $n_\tau$ plans whose loss relative to the optimal conditional value is at most $\tau$. Draw $n_{\text{draw}}$ candidates independently and uniformly with replacement, and retain the best evaluated plan or the incumbent. If $\Delta_U$ bounds the full feasible payoff range, then

$$\mathbb{E}[\text{regret} \mid z] \leq \tau + \Delta_U (1 - n_\tau/N)^{n_{\text{draw}}}. \tag{A6}$$

A draw misses every near-optimal candidate with probability $1 - n_\tau/N$. Independence gives the power in equation (A6). When at least one such candidate is found, loss is at most $\tau$; otherwise it is at most $\Delta_U$. This proves the bound. It is a uniform-random-search result, not a guarantee for a dependent genetic search. The number and distribution of useful alternatives matter alongside the total count.

For Proposition 2, choose a true conditional maximizer $x^*$ and insert the estimated values:

$$\begin{aligned} m(x^*) - m(\hat{x}) = {} & [m(x^*) - \hat{m}(x^*)] \\ & + [\hat{m}(x^*) - \hat{m}(\hat{x})] \\ & + [\hat{m}(\hat{x}) - m(\hat{x})] \\ & \leq \varepsilon + \delta + \varepsilon. \end{aligned} \tag{A7}$$

The first and third terms use uniform valuation accuracy, and the middle term uses the stated optimization guarantee. The result concerns conditional expected decision value; it is not a pointwise guarantee about realized outcomes in every state.

### A.5 Organizational design and decision capability

Suppose the simple design yields $v_S \geq 0$ and switching to the coordinated-project design costs $C_{\text{switch}} \geq 0$. The values are

$$J_C = \max\{0, \Phi\} - C_{\text{switch}}, \qquad J_S = v_S. \tag{A8}$$

The interdependent design is strictly preferable exactly when $\Phi > v_S + C_{\text{switch}}$. On the interior precision branch, this becomes

$$\frac{V^2}{8\kappa} > c + F_E + v_S + C_{\text{switch}}, \qquad \kappa \geq V/2. \tag{A9}$$

An integrated closed form for the design choice is

$$\begin{aligned} J_{\text{choose}} = \max\{v_S, \frac{V}{2}\min\{\frac{V}{2\kappa}, 1\} - c \\ - \frac{\kappa}{2}(\min\{\frac{V}{2\kappa}, 1\})^2 \\ - F_E - C_{\text{switch}}\}. \end{aligned} \tag{A10}$$

The unused option of entering the interdependent design without investing has value $-C_{\text{switch}}$, which cannot improve on $v_S \geq 0$. This explains its omission from the maximum. The extension shows how decision costs can affect design choice under the stated payoffs; it does not imply that analytical capability makes every interdependent structure preferable.

## Appendix B Computational specification

### B.1 Hypothetical educational response functions

For the practices in Table 2, the generating equations are

$$\begin{aligned} a(x,\theta) = {} & 0.08 - 0.018x_1 + 0.015x_2 - 0.012x_3 \\ & -0.010x_4 - 0.012x_5 - 0.010x_6 \\ & -0.020\theta_1 x_1 x_2, \end{aligned} \tag{B1}$$

$$\begin{aligned} d(x,\theta) = {} & 0.65 + 0.040x_1 + 0.060x_2 + 0.025x_3 \\ & -0.015x_4 + 0.020x_5 + 0.005x_6 \\ & +0.030\theta_1 x_1 x_2, \end{aligned} \tag{B2}$$

$$\begin{aligned} q(x,\theta) = {} & 0.70 - 0.005x_1 + 0.010x_2 + 0.050x_3 \\ & -0.010x_4 + 0.035x_5 + 0.005x_6 \\ & +0.025\theta_2 x_3 x_5. \end{aligned} \tag{B3}$$

The independent interaction states each equal minus one or plus one with probability one half. Coefficients encode trade-offs and interactions while keeping outcomes within [0,1] on the tested feasible sets. They are assumptions, not fitted estimates or externally established educational effect sizes. Substitution into equation (9) gives

$$\begin{aligned} U(x,\theta) = {} & 5.3x_1 + 5.5x_2 + 8.7x_3 - 1.5x_4 + 6.7x_5 + 2x_6 \\ & +5\theta_1 x_1 x_2 + 2.5\theta_2 x_3 x_5 - \Gamma(x)/10{,}000. \end{aligned} \tag{B4}$$

The code enumerates $\{0,0.5,1\}^6$ in lexicographic order and filters by cumulative transition cost. The incumbent is the first retained configuration. The four states are ordered $(-1,-1), (-1,1), (1,-1), (1,1)$. The exact scalar maxima are 12.25, 17.25, 14.65, and 19.65. The first two states select $(0,0.5,1,0,1,0.5)$ at EUR 44,000; the last two select $(0.5,0.5,1,0,1,0)$ at EUR 49,000. Their average value is 15.95. The best uninformed value is 15.90.

### B.2 Information and random search experiment

The information experiment uses NumPy's default PCG64 generator with seed 20260926 and 10,000 replications. It draws a 10,000 by 50 array of nonincumbent feasible indices with replacement, using the first ten entries for the lower budget and all fifty for the higher budget. The incumbent is included separately. Hidden-state selection maximizes the average payoff over four states. Revealed-state selection maximizes separately in each state before averaging the results. Ties use the first maximizer.

The Monte Carlo standard error is the sample standard deviation across replications divided by the square root of their number. The cross-difference is computed within each replication before averaging. Sensitivity analysis uses budgets EUR 25,000, EUR 50,000, and EUR 75,000, with absence and exit weights independently set to 0.5, 1, and 2 while the completion weight remains one. Each setting resets the generator to 20260927 and uses 2,000 replications. This reuse supports common-random-number comparisons; settings are not independent empirical samples.

### B.3 Genetic search and Pareto evaluation

The reference search encodes each practice as integer 0, 1, or 2. These integers are category indices corresponding to implementation levels 0, 0.5, and 1. Uniform initialization and the random-search comparator draw category vectors independently, rejecting over-budget vectors before an outcome evaluation. The incumbent is included once initially. Genetic population size is 20, initialized with the incumbent and 19 feasible draws.

Binary tournaments prefer lower nondomination rank and then higher crowding distance, resolving remaining ties randomly. Uniform crossover occurs with probability 0.9, and each coordinate independently selects a parent's category with probability one half. Mutation changes each coordinate with probability $1/6$, selecting either other category with equal probability. Infeasible offspring are rejected. Each generation produces 20 accepted offspring. The next population is selected from parents and offspring by rank and crowding, with random tie breaking. These are specified discrete variation operators around the NSGA-II selection mechanism.

Search is observed after 60 and 200 total candidate evaluations, with lower-budget evaluations nested inside the higher-budget run. Duplicate candidates consume an evaluation. The reported archive contains nondominated configurations among all evaluated candidates, not only the final population. The four objectives are negative absence, completion, exit index, and negative euro cost. Crowding uses objective-specific ranges; numerical dominance comparisons use tolerance $10^{-10}$.

There are 50 runs per state. Genetic seeds are $20260928 + 1000h + r$, where $h \in \{0,1,2,3\}$ indexes the stated state order and $r \in \{0, \dots, 49\}$ indexes the replication. Random-search seeds add 100000 to that expression. Exact frontier membership is used only for assessment, not for generating candidates or selecting parents. The code checks the vectorized frontier against an independent pairwise implementation and verifies that each exact scalar optimum belongs to the corresponding frontier.

Frontier recovery compares configuration identifiers with the exact set. It is a stringent exact-recovery measure and does not capture every aspect of approximate frontier quality. The scalar gap supplies a second, preference-specific measure. The small instance, fixed operators, and two budgets do not establish performance scaling or algorithmic superiority. The exported per-run records permit alternative summaries without hiding unsuccessful searches.

### B.4 Action compilation and reproduction

The selected example is the exact optimum in state (1,1), so the implementation illustration does not depend on a fortunate heuristic run. A compiler traverses the required adjacent transitions for each changed practice, copies the corresponding cost, hours, and nominal duration, and records within-practice predecessors. It verifies that the six transitions sum to EUR 49,000 and 450 hours and that their terminal vector equals the evaluated plan. Owner roles and acceptance checks are explicitly proposed fields.

The package contains the information and search experiment, the Pareto and action demonstration, mathematical verification scripts, outputs, and instructions. The scripts write exact frontiers, per-run search results, search summaries, and the selected action specification. No confidential team records or observed KPI rows are required for reproduction. Exported simulated configurations are model evaluations, not an enlarged sample of observed teams.

# Appendix C Notation and the mathematical bridge

## C.1 Shared notation

**Table C1 Main symbols and their scope**

| Symbol | Meaning |
|---|---|
| $j, n_P, s_j$ | Practice index, number of practices, and terminal state of a practice |
| $g, s^0, x^0$ | Organizational design, incumbent state profile, and incumbent plan |
| $x = (s, \rho)$ | Plan with terminal profile and a permitted transition path; the simulation's fixed paths allow $x$ to denote its level vector |
| $\mathcal{X}_B, B, \Gamma$ | Feasible plans, euro budget, and euro expenditure |
| $\theta, z, S$ | Uncertain circumstances, received information, and signal experiment |
| $y_k, w_k, T$ | Favorably oriented outcome change, valuation weight, and implementation charge in common value units |
| $U, W, J(g)$ | Plan value, attainable expected value before decision charges, and optimized value after those charges |
| $b, \Pi_b, \pi$ | Evaluation capability, admissible procedure class, and a decision procedure |
| $C_I, C_E$ | Information and evaluation charges; these are distinct from practice cancellation $C_i$ |
| $E_i, M_i, D_i$ | Nonnegative practice intensity, total absolute valued effect per unit, and signed net valued effect per unit |
| $C_i, \sigma_i, H$ | Cancellation fraction, sign of the net contribution, and retained joint effects |
| $L_I, L_E, \eta$ | Information loss, evaluation loss, and normalized value-based visibility |
| $\mathbf{f}, \mathcal{P}$ | Vector of favorably oriented objectives and the conditional Pareto set |
| $\varepsilon, \delta$ | Uniform valuation-error bound and estimated-objective optimization gap |
| $V, c, \kappa, F_E, e$ | Joint benefit magnitude, cost per change, information-cost parameter, capability charge, and signal precision in the binary example |
| $v_S, C_{\text{switch}}$ | Value of the simple design and design-transition charge |
| $a, d, q$ | Simulated absence rate, completion rate, and exit-performance index |

## C.2 Decomposing the education objective

For equation (9), use $y_1 = 0.08 - a$, $y_2 = d - 0.65$, and $y_3 = q - 0.70$, with all three valuation weights equal to 100. Implementation intensity is $E_i = x_i$, and the charge is $T = \Gamma(x)/10{,}000$. Table C2 gives the practice-level decomposition of the main effects in equation (B4).

**Table C2 Main practice effects under the illustrative valuations**

| Practice | Signed effect $D_i$ | Magnitude $M_i$ | Cancellation $C_i$ | Sign $\sigma_i$ |
|---|---|---|---|---|
| Q1 Attendance recording | 5.3 | 6.3 | 0.1587 | +1 |
| Q3 Absence follow-up | 5.5 | 8.5 | 0.3529 | +1 |
| Q12 Test-result processing | 8.7 | 8.7 | 0 | +1 |
| Q19 Daily work locations | −1.5 | 3.5 | 0.5714 | −1 |
| Q31 ICT assistance | 6.7 | 6.7 | 0 | +1 |
| Q34 Other administration | 2.0 | 2.0 | 0 | +1 |

For Q1 the valued effects are $(1.8, 4.0, -0.5)$, giving $D_1 = 5.3$, $M_1 = 6.3$, and $C_1 = 10/63$. Thus $\sigma_1 E_1 M_1(1 - C_1) = 5.3x_1$. The corresponding exact cancellation fractions for Q3 and Q19 are $6/17$ and $4/7$. The joint term is $H = 5\theta_1 x_1 x_2 + 2.5\theta_2 x_3 x_5$. Substitution into equation (2) reconstructs equation (B4). The cancellation table uses rounded display values only; verification uses the unrounded quantities.

This equality connects the articles without identifying an additional empirical relationship. It neither proves that cancellation can be removed nor estimates the gains from doing so. The relevant comparison remains between feasible plans under the same horizon, valuations, constraints, and cost accounting.

## Data and code availability

Scripts, scenario parameters, selected indicative transition inputs, numerical outputs, and instructions accompany the article. Reproduction requires no observed team records. The supplied team workbook and respondent records are excluded; no public availability or third-party reuse authorization is asserted.

## Declaration of generative AI assistance

Generative AI assisted with drafting, editing, mathematical exposition, and code preparation and checking. The author remains responsible for verifying all arguments, sources, calculations, disclosures, and the final manuscript before submission.

## Competing interests

The author founded Transparency Lab, owner of PRAIORITIZE. This commercial connection is relevant to the proposed application. The theoretical results and simulations are not an independent evaluation of the product.

## References

Bertsimas, D., and Kallus, N. (2020). From predictive to prescriptive analytics. *Management Science, 66*(3), 1025–1044. https://doi.org/10.1287/mnsc.2018.3253

Blackwell, D. (1953). Equivalent comparisons of experiments. *The Annals of Mathematical Statistics, 24*(2), 265–272. https://doi.org/10.1214/aoms/1177729032

Cawley, G. C., and Talbot, N. L. C. (2010). On over-fitting in model selection and subsequent selection bias in performance evaluation. *Journal of Machine Learning Research, 11*, 2079–2107. https://jmlr.org/papers/v11/cawley10a.html

Deb, K., Pratap, A., Agarwal, S., and Meyarivan, T. (2002). A fast and elitist multiobjective genetic algorithm NSGA-II. *IEEE Transactions on Evolutionary Computation, 6*(2), 182–197. https://doi.org/10.1109/4235.996017

Dessein, W., Galeotti, A., and Santos, T. (2016). Rational inattention and organizational focus. *American Economic Review, 106*(6), 1522–1536. https://doi.org/10.1257/aer.20140741

Elmachtoub, A. N., and Grigas, P. (2022). Smart "Predict, then Optimize". *Management Science, 68*(1), 9–26. https://doi.org/10.1287/mnsc.2020.3922

Garicano, L. (2000). Hierarchies and the organization of knowledge in production. *Journal of Political Economy, 108*(5), 874–904. https://doi.org/10.1086/317671

Gershman, S. J., Horvitz, E. J., and Tenenbaum, J. B. (2015). Computational rationality: A converging paradigm for intelligence in brains, minds, and machines. *Science, 349*(6245), 273–278. https://doi.org/10.1126/science.aac6076

Karimi, A.-H., Schölkopf, B., and Valera, I. (2021). Algorithmic recourse: From counterfactual explanations to interventions. *Proceedings of the 2021 ACM Conference on Fairness, Accountability, and Transparency*, 353–362. https://doi.org/10.1145/3442188.3445899

Levinthal, D. A. (1997). Adaptation on rugged landscapes. *Management Science, 43*(7), 934–950. https://doi.org/10.1287/mnsc.43.7.934

Morris, T. P., White, I. R., and Crowther, M. J. (2019). Using simulation studies to evaluate statistical methods. *Statistics in Medicine, 38*(11), 2074–2102. https://doi.org/10.1002/sim.8086

Oprea, R. (2020). What makes a rule complex? *American Economic Review, 110*(12), 3913–3951. https://doi.org/10.1257/aer.20191717

Simon, H. A. (1955). A behavioral model of rational choice. *The Quarterly Journal of Economics, 69*(1), 99–118. https://doi.org/10.2307/1884852

Sims, C. A. (2003). Implications of rational inattention. *Journal of Monetary Economics, 50*(3), 665–690. https://doi.org/10.1016/S0304-3932(03)00029-1

Smith, J. E., and Winkler, R. L. (2006). The optimizer's curse: Skepticism and postdecision surprise in decision analysis. *Management Science, 52*(3), 311–322. https://doi.org/10.1287/mnsc.1050.0451

Ustun, B., Spangher, A., and Liu, Y. (2019). Actionable recourse in linear classification. *Proceedings of the Conference on Fairness, Accountability, and Transparency*, 10–19. https://doi.org/10.1145/3287560.3287566

Van de Poll, J. (2021). An alternative to the Likert scale when polling employees. *The International Journal of Business & Management, 9*(5), 239–244. https://doi.org/10.24940/theijbm/2021/v9/i5/BM2105-044

Van de Poll, J. (2026a). *Unpriced internal externalities and structural inefficiency in organizations*. Revised working paper. Transparency Lab Research Institute.

Van de Poll, J. (2026b). *Organizational layering and the persistence of internal externalities*. Revised working paper. Transparency Lab Research Institute.